\documentclass[apjl]{emulateapj}
\journalinfo{To appear in The Astronomical Journal}

\def\kms{\ifmmode{\rm km\thinspace s^{-1}}\else km\thinspace s$^{-1}$\fi}
\def\hr{HR~6046}
\def\hip{{\it Hipparcos\/}}

\shortauthors{Torres}
\shorttitle{\hr}

\begin{document}

\title{HIPPARCOS astrometric orbit and evolutionary status of
\hr}

\author{Guillermo Torres}

\affil{Harvard-Smithsonian Center for Astrophysics, 60 Garden St.,
Cambridge, MA 02138}

\email{gtorres@cfa.harvard.edu}

\begin{abstract} 

The previously known, 6-yr spectroscopic binary \hr\ has been
speculated in the past to contain a compact object as the secondary. A
recent study has re-determined the orbit with great accuracy, and
shown that the companion is an evolved but otherwise normal star of
nearly identical mass as the primary, which is also a giant. The
binary motion was detected by the \hip\ mission but was not properly
accounted for in the published astrometric solution. Here we use the
\hip\ intermediate data in combination with the spectroscopic results
to revise that solution and establish the orbital inclination angle
for the first time, and with it the absolute masses $M_{\rm A} =
1.38_{-0.03}^{+0.09}$~M$_{\sun}$ and $M_{\rm B} =
1.36_{-0.02}^{+0.07}$~M$_{\sun}$. Aided by other constraints, we
investigate the evolutionary status and confirm that the primary star
is approaching the tip of the red-giant branch, while the secondary is
beginning its first ascent.
	
\end{abstract}


\section{Introduction}
\label{sec:introduction}

Since the publication of the first preliminary orbits in the 1930's by
\cite{Christie:34, Christie:36}, the bright giant star \hr\
(HD~145849, HIP~79358, $\alpha = 16^{\rm h} 11^{\rm m} 48\fs05$,
$\delta = +36\arcdeg 25\arcmin 30\farcs3$, J2000.0; spectral type
\ion{K3}{2}, $V = 5.63$) has been known as a highly eccentric
$\sim$6-yr period single-lined spectroscopic binary of particular
interest. The relatively large minimum mass inferred early on for the
companion ($> 3$~M$_{\sun}$), along with the fact that it was not
visible to early observers, led to the speculation that the secondary
was a collapsed star \citep{Trimble:69}, more commonly referred to
nowadays as a black hole.  Very recently \cite{Scarfe:07} have
dispelled this notion by detecting the secondary spectroscopically and
showing that it is merely an evolved late-type star, with nothing
particularly out of the ordinary in its properties other than the fact
that it is faint. On the basis of extensive radial velocity
measurements carried out over more than 26 years these authors
presented an accurate double-lined orbit for the system yielding
minimum masses considerably smaller than previously thought. They also
found the components to be nearly identical in mass to within about
1\%, even though they are substantially different in brightness
($\Delta V \approx 3$). They relied on the \hip\ parallax of the
system \citep{Perryman:97} to construct a plausible photometric model
of the stars (individual magnitudes and colors), but cautioned that
the distance could be vitiated by unmodeled photocentric motion.

As described by \cite{Scarfe:07}, \hr\ has never been spatially
resolved despite repeated attempts over the years using speckle
interferometry and other techniques. However, \cite{Jancart:05} were
able to obtain an estimate of the inclination angle by reconsidering
the \hip\ Intermediate Astrometric Data in conjunction with the
spectroscopic orbit of \cite{Christie:36} to account for the binary
motion. With this the absolute masses of the components can in
principle be obtained, although the resulting values are somewhat high
for late-type giants ($\sim$3.6 M$_{\sun}$) possibly because of the
use of outdated spectroscopic elements.

The main motivation for the present work is to revisit the \hip\
astrometric solution in the light of the accurate orbit of
\cite{Scarfe:07}, and to examine the effect on the published
parallax. Additionally, we investigate the evolutionary status of the
system with our newly determined masses, aided by current stellar
evolution models. We report also a spectroscopic determination of the
effective temperature of the primary star that supports the general
picture outlined by other constraints in showing the evolved state of
the binary.

\section{HIPPARCOS observations and revised astrometric solution}
\label{sec:orbit}

Given the 6-yr orbital period of \hr, which is of the same order as
the duration of the \hip\ mission (slightly more than 3 yr), it may be
expected that the orbital motion would be detectable in the satellite
measurements, and that if not properly accounted for, it could bias
either the trigonometric parallax, the proper motions, or both.
Indeed, proof that a signal of this nature is detectable is given by
the fact that the \hip\ team found it necessary to include
acceleration terms in the astrometric solution, representing the first
derivatives of the proper motions. These terms, $d\mu_{\alpha}^*/dt =
-9.08 \pm 1.23$ mas~yr$^{-2}$ and $d\mu_{\delta}/dt = -6.98 \pm 1.53$
mas~yr$^{-2}$, are statistically significant. It is unclear why a full
orbital model was not applied originally to this system, since the
binary nature of the object has been known for a long time. In any
case, this has been done more recently by \cite{Jancart:05}, as
mentioned earlier.

A total of 78 astrometric measurements were made by \hip\ from 1989
December to 1993 February, covering about 53\% of an orbital cycle.
Each measurement consisted of a one-dimensional position (`abscissa',
$v$) along a great circle representing the scanning direction of the
satellite, tied to an absolute frame of reference known as the
International Celestial Reference System (ICRS). The measurements
available for our analysis are published in the form of `abscissa
residuals' ($\Delta v$; see Table~\ref{tab:abscissae}), which are the
residuals from the standard five-parameter solution reported in the
Catalogue \citep{Perryman:97}.  The five standard parameters are the
position ($\alpha_0^*$, $\delta_0$) and proper motion components
($\mu_{\alpha}^*$, $\mu_{\delta}$) of the barycenter at the reference
epoch 1991.25\footnote{Following the practice in the \hip\ Catalogue
we define $\alpha^* \equiv \alpha \cos\delta$ and $\mu_{\alpha}^*
\equiv \mu_{\alpha} \cos\delta$.}, and the trigonometric parallax,
$\pi_{\rm t}$. The nominal errors of these measurements have a median
value of 2.5 mas.

We incorporated orbital motion into a new astrometric solution
following the formalism of \cite{vanLeeuwen:98}, \cite{Pourbaix:00},
and \cite{Jancart:05}, including the correlations between measurements
from the two independent data reduction consortia that processed the
original \hip\ observations \citep[NDAC and FAST;][]{Perryman:97}.
Details of this modeling along with another example of an application
may also be seen in \cite{Torres:07b}. In the most general case the
adjustable quantities of the fit (a total of 12) are the corrections
to the five standard \hip\ parameters ($\Delta\alpha^*$,
$\Delta\delta$, $\Delta\mu_{\alpha}^*$, $\Delta\mu_{\delta}$, and
$\Delta\pi_{\rm t}$), and the seven usual elements of the binary
orbit: the period $P$, the semimajor axis of the photocenter $a_{\rm
phot}$, the eccentricity $e$, the inclination angle $i$, the longitude
of periastron for the primary $\omega_{\rm A}$, the position angle of
the ascending node $\Omega$ (equinox J2000.0), and the time of
periastron passage $T$. For \hr\ the \hip\ measurements do not
constrain $P$, $e$, $\omega_{\rm A}$, or $T$ sufficiently well, so
those elements were held fixed at their spectroscopic values as
reported by \cite{Scarfe:07}. We solved for the remaining 8 parameters
simultaneously using standard non-linear least-squares techniques
\citep[][p.\ 650]{Press:92}. The reduced $\chi^2$ of the solution is
0.9895, indicating that the internal errors of the abscissa residuals
are realistic.

We list the results in Table~\ref{tab:elements}, where they are
compared with those obtained by \cite{Jancart:05}. Significant
differences are seen in the semimajor axis and inclination angle,
which reflect the spectroscopic constraints used in each case.
Jancart's value of $a_{\rm phot}$ is nearly twice as large as ours,
and while their inclination angle is substantially smaller than
90\arcdeg, we obtain an orientation that is essentially edge-on. The
smaller $i$ value of \cite{Jancart:05} would lead to much larger
absolute masses for the stars when combined with the minimum masses of
\cite{Scarfe:07}, as noted in \S\ref{sec:introduction}.  There is
little doubt that the recent orbital elements by \cite{Scarfe:07}
supersede the provisional values of \cite{Christie:36} adopted by
Jancart, which suggests our solution should be closer to the
truth.\footnote{A preliminary version of the \cite{Scarfe:07} work had
been published in 2004 \citep{Scarfe:04}, but was perhaps not yet
available to Jancart and collaborators.} An external check on our
results is provided by the absolute proper motions we obtain, listed
in the bottom section of Table~\ref{tab:elements}. These are different
from the values reported in the \hip\ Catalogue, which are likely to
be affected also by the unmodeled orbital motion. They agree well,
however, with the motions given in the Tycho-2 Catalogue
\citep{Hog:00}, as shown in Table~\ref{tab:pm}. The latter are based
on the position from the Tycho experiment aboard the \hip\ satellite
(epoch $\sim$1991.25) combined with ground base catalog positions
going back nearly a century in some cases.  This long baseline tends
to average out any orbital motion that has a period of a few years, as
in the case of \hr, and therefore yields a more accurate estimate than
{\it Hipparcos\/}.  Finally, we note that our revised \hip\ solution
did not change the value of the parallax significantly, which was one
of the concerns of \cite{Scarfe:07}. This parallax corresponds to a
distance of 178 pc.

\begin{figure}
\vskip -0.3in
\epsscale{1.3}
{\hskip -0.2in \plotone{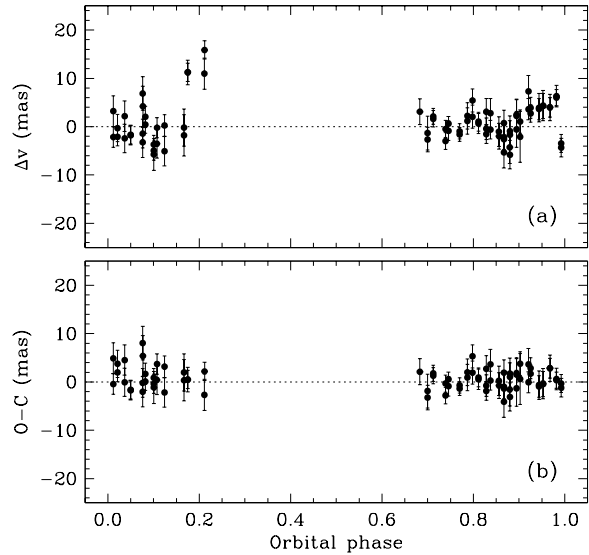}}
\vskip -0.3in 
\figcaption[]{(a) Abscissa residuals of \hr\ from the \hip\ mission as
a function of orbital phase. These are the residuals from the standard
5-parameter solution as published in the Catalogue
\citep{Perryman:97}.  The systematic patterns are due to the unmodeled
motion of the center of light. (b) $O\!-\!C$ residuals of the \hip\
measurements from our new 12-parameter fit that accounts for the
orbital motion.\label{fig:hipresid}}
\end{figure}

As a way of visualizing the effect of accounting for the orbital
motion in the \hip\ solution, we show in Figure~\ref{fig:hipresid}a
the abscissa residuals $\Delta v$ resulting from the standard
5-parameter solution as a function of orbital phase. Some systematic
patterns are apparent, but they largely disappear after the orbital
motion is incorporated into the model. This is shown by the $O\!-\!C$
residuals from the 12-parameter fit in Figure~\ref{fig:hipresid}b.

\begin{figure} 
\vskip -0.6in
\epsscale{1.3} 
{\hskip -0.2in \plotone{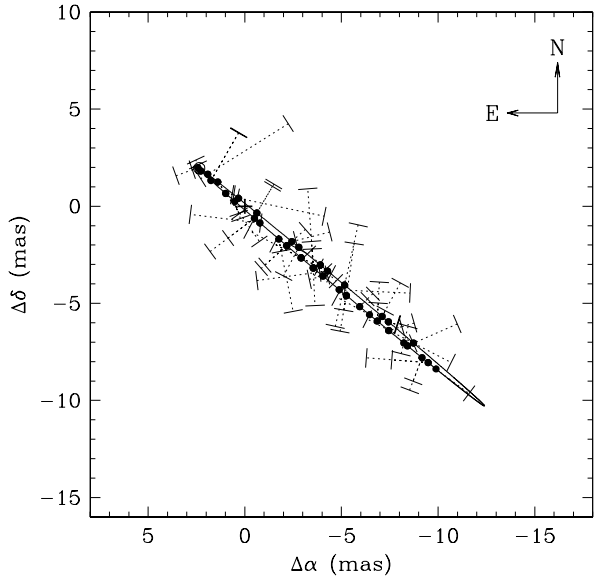}}
\figcaption[]{Motion of the photocenter of \hr\ relative to the center
of mass of the binary (indicated by the plus sign) as seen by
\hip. The one-dimensional abscissa residuals are shown schematically
with a filled circle at the predicted location, dotted lines
representing the scanning direction of the satellite, and short
perpendicular line segments indicating the undetermined location of
the measurement on that line (see text). The length of the dotted
lines represents the magnitude of the $O\!-\!C$ residual from the
computed location.  Three measurements with residuals larger than 5
mas were omitted for clarity. Also indicated on the plot is the
location of periastron (open circle near the
top).\label{fig:hiporbit}}
\end{figure}

\begin{figure} 
\vskip -0.6in
\epsscale{1.3} 
{\hskip -0.2in \plotone{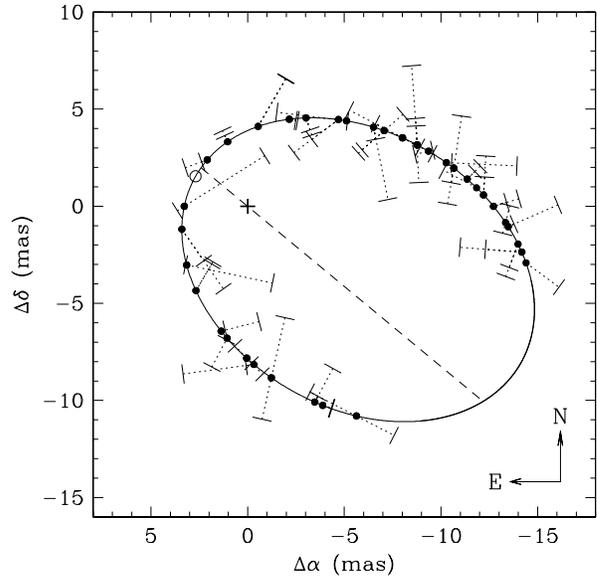}}
\figcaption[]{Same as Figure~\ref{fig:hiporbit}, except that the orbit
has been de-projected to appear as if it were viewed exactly face-on.
The line of nodes is indicated with the dashed line. Motion on the
plane of the sky in this figure is direct (counterclockwise).
\label{fig:deproject}}
\end{figure}

The projection of the photocentric orbit on the plane of the sky along
with a schematic representation the \hip\ measurements is seen in
Figure~\ref{fig:hiporbit}, where the axes are parallel to the right
ascension and declination directions.  Because these measurements are
one-dimensional in nature, their exact location on the plane of the
sky cannot be shown graphically. The filled circles represent the
predicted location on the computed orbit (see also
Figure~\ref{fig:deproject}).  The dotted lines connected to each
filled circle indicate the scanning direction of the \hip\ satellite
for each measurement, and show which side of the orbit the residual is
on. The length of each dotted line represents the magnitude of the
$O\!-\!C$ residual.\footnote{The ``$O\!-\!C$ residuals'' are not to be
confused with the ``abscissa residuals'' $\Delta v$, which we refer to
loosely here as \hip\ ``observations'' or ``measurements''. As
indicated earlier, the abscissa residuals are in fact residuals from
the standard 5-parameter fit reported in the \hip\ Catalogue, whereas
the $O\!-\!C$ residuals (or simply ``residuals'') are the difference
between the abscissa residuals and the computed position of the star
from a model that incorporates orbital elements.}  The short line
segments at the end of and perpendicular to the dotted lines indicate
the direction along which the actual observation lies, although the
precise location is undetermined. Occasionally more than one
measurement was taken along the same scanning direction, in which case
two or more short line segments appear on the same dotted lines. The
orbit is formally clockwise (retrograde), although the orientation is
so close to edge-on ($i = 91\fdg4 \pm 9\fdg7$) that the motion on the
plane of the sky may well be direct.

\begin{figure}
\vskip -0.6in
\epsscale{1.3}
{\hskip -0.2in \plotone{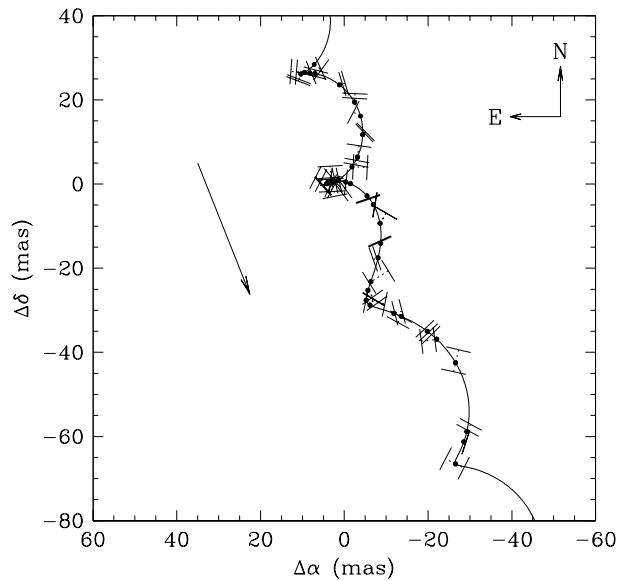}}
\figcaption[]{Path of the center of light of \hr\ on the plane of the
sky, along with the \hip\ observations (abscissa residuals)
represented as in Figure~\ref{fig:hiporbit} and
Figure~\ref{fig:deproject}. The irregular motion is the result of the
combined effects of the annual parallax, proper motion, and orbital
motion according to the solution described in the text.  The arrow
indicates the direction and magnitude of the annual proper
motion.\label{fig:hippath}}
\end{figure}

The phase coverage of the \hip\ observations is seen more clearly in
Figure~\ref{fig:deproject}, in which we have de-projected the orbit
and represented it as if it were viewed with an inclination angle of
0\arcdeg. The measurements happen to cover the periastron passage of
1991.83 (indicated with an open circle), but not apastron.  The path
of \hr\ on the plane of the sky as seen by \hip\ is shown in
Figure~\ref{fig:hippath}.  The irregular pattern is the result of the
combination of annual proper motion (indicated by the arrow),
parallactic motion, and orbital motion.

The detection of the photocentric motion by \hip\ offers an
independent way of estimating the brightness difference between the
stars, in the passband of the satellite ($H_p$ band), from the
relation between the semimajor axis of the photocentric orbit and that
of the relative orbit in angular units \citep[e.g.,][]{vandeKamp:67}:
$a_{\rm phot} = a_{\rm rel}(B-\beta)$. Here $a_{\rm rel}$ can be
obtained from the spectroscopic minimum masses, the inclination angle,
and the parallax through Kepler's Third Law. $B$ represents the mass
fraction $M_{\rm B}/(M_{\rm A}+M_{\rm B})$, and $\beta$ is the light
fraction given by $L_{\rm B}/(L_{\rm A}+L_{\rm B}) = (1+10^{0.4\Delta
H_p})^{-1}$. The magnitude difference is then
\begin{equation}
\Delta H_p = 2.5 \log \left[ \left({q\over 1+q}-{a_{\rm phot}\over
a_{\rm rel}}\right)^{-1} - 1 \right]~,
\end{equation}
in which $q \equiv M_{\rm B}/M_{\rm A}$.  We obtain $\Delta H_p = 2.1
\pm 0.6$ mag, somewhat smaller but still broadly consistent with the
values in the Johnson $B$ and $V$ bands reported by \cite{Scarfe:07}
(see below), considering the difference in the passbands.

\section{Evolutionary status}
\label{sec:evolution}

From the analysis of their high-resolution spectra of \hr\
\cite{Scarfe:07} estimated the spectral types and luminosity classes
of the components to be \ion{K3}{2} and \ion{K0}{4}, and the
brightness difference to be $\Delta B \approx 2.5$ mag and $\Delta V
\approx 3.0$ mag. They then made use of the \hip\ parallax (which we
now know is substantially correct) and the combined system brightness
to infer absolute visual magnitudes of $M_V = -0.68$ for the primary
and $M_V = 2.32$ for the secondary, ignoring extinction.  $U\!-\!B$
and $B\!-\!V$ color indices were computed by making use of standard
tabulated values for spectral types and luminosity classes close to
what they derived, along with the magnitude differences. The
photometry synthesized in this way provides a very good match to the
observed $B\!-\!V$ color of the system, and enabled \cite{Scarfe:07}
to claim that both stars are evolved, with the primary being near the
tip of the giant branch. They also pointed out that evolutionary
models by \cite{Pols:98} allow the brightness difference to be as
large as observed in \hr\ for component masses that differ by as
little as they do in this system ($\sim$1\%), at least for masses in
the most probable range for \hr, which they estimated to be 2.0--3.5
M$_{\sun}$. This provided a natural explanation for a set of
properties that had intrigued earlier investigators.

The minimum masses from the spectroscopic work of \cite{Scarfe:07}
combined with our inclination angle give absolute masses for \hr\ of
$M_{\rm A} = 1.38_{-0.03}^{+0.09}$~M$_{\sun}$ and $M_{\rm B} =
1.36_{-0.02}^{+0.07}$~M$_{\sun}$ (Table~\ref{tab:elements}), in which
the uncertainties are currently dominated by the error in $i$. These
masses are considerably smaller than assumed by those authors. We have
therefore re-examined the consistency with the evolutionary models by
making use of these values along with slightly revised absolute
magnitudes for the stars based on our new parallax. Given a distance
approaching 200~pc, a small amount of extinction would not be
unexpected for the system, and in fact the dust maps of
\cite{Schlegel:98} indicate a total reddening in the direction of \hr\
of $E(B\!-\!V) \sim 0.03$ mag. We tentatively adopt $E(B\!-\!V) \sim
0.02 \pm 0.01$ mag, which corresponds to $A_V \sim 0.06 \pm 0.03$ mag.
With this adjustment to the apparent magnitude $V = 5.63 \pm 0.01$
\citep{Haggkvist:87}, and taking into account the magnitude difference
from \cite{Scarfe:07} (to which we assign, somewhat arbitrarily, an
uncertainty of 0.3 mag), we infer individual magnitudes of $M_V^{\rm
A} = -0.62 \pm 0.20$ and $M_V^{\rm B} = 2.38 \pm 0.34$. These are not
changed much from the original estimates of \cite{Scarfe:07}.  The
$B\!-\!V$ colors proposed by them rely on an external tabulation and
are quite sensitive to the luminosity class adopted, so we have
preferred to proceed without them here, although they may well be
accurate. The models by \cite{Girardi:00} are a convenient choice for
a comparison with the observations, since they reach the late
evolutionary stages needed for the primary and are tabulated for a
wide range of chemical compositions and ages. They also incorporate
mass loss due to winds in the giant phase, which turns out to be only
a 1--2\% effect for \hr.

\begin{figure}
\vskip -0.4in
\epsscale{1.4}
{\hskip -0.35in \plotone{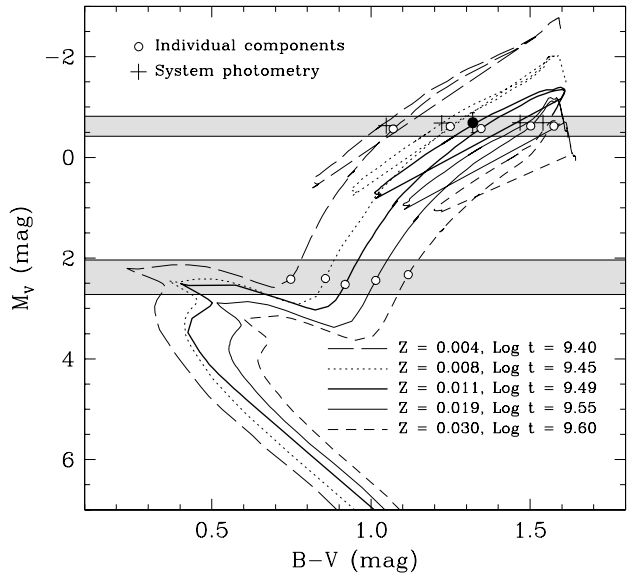}}
\vskip -0.25in
\figcaption[]{Model isochrones from \cite{Girardi:00} that match the
measured absolute magnitudes (grey areas) and current masses of \hr\
within the observational errors, shown for a range of metallicities
and ages as labeled. Open circles on each isochrone indicate the
expected location of the components, and the plus signs represent the
predicted system magnitude and color. The system magnitude and color
actually measured are represented by the filled circle and error
bar. The model drawn with the heavy line satisfies the additional
constraint given by the observed $B\!-\!V$ color of the system.
\label{fig:isochrones}}
\end{figure}

In Figure~\ref{fig:isochrones} we show several isochrones for
different metallicities from this series of models that produce the
best simultaneous match to the absolute magnitudes of both components
at their current masses (accounting for mass loss). In each case
(except for the middle isochrone) the age is the one providing the
best agreement among the values tabulated by \cite{Girardi:00}, which
come in steps of 0.05 in $\log t$.  The location of the stars on these
models is indicated with open circles. The combined color and
magnitude predicted for the system in each case is shown by the plus
signs, and the filled circle represents the measured values of $M_V$
and $B\!-\!V$ for \hr. The system color we have adopted is $B\!-\!V =
1.34 \pm 0.01$ \citep{Haggkvist:87}, which we have then dereddened as
indicated above.  Properties of the stars inferred from these models
are listed in Table~\ref{tab:models}. If we now impose the additional
requirement that the isochrones match the measured system color within
its uncertainty, a very good correspondence may be obtained by
interpolation for a metallicity of $Z = 0.011 \pm 0.002$
(corresponding to [Fe/H] $= -0.24 \pm 0.07$ for the assumed $Z_{\sun}
= 0.019$ in these models) and a logarithmic age of $\log t =
9.49_{-0.09}^{+0.04}$ (or $3.1_{-0.6}^{+0.3}$~Gyr). This model is
shown with the heavy line in Figure~\ref{fig:isochrones}, and
indicates that the primary is indeed approaching the tip of the giant
branch, whereas the secondary is beginning its first ascent of the
giant branch. We infer from this best-fitting isochrone that the
primary star has a radius of $30.9_{-1.1}^{+3.6}$~R$_{\sun}$, a
surface gravity of $\log g = 1.60_{-0.08}^{+0.04}$, and an effective
temperature ($T_{\rm eff}$) of $4211_{-23}^{+16}$~K, and the secondary
has $R = 4.20_{-0.33}^{+0.27}$~R$_{\sun}$, $\log g =
3.33_{-0.04}^{+0.07}$, and $T_{\rm eff} = 5010_{-47}^{+79}$~K. These
uncertainties account for all observational errors but exclude
possible systematics in the models.  Other properties of the stars for
this metallicity are listed in Table~\ref{tab:models}.

\subsection{Additional spectroscopic constraints}

With the goal of providing a check on the temperature of the primary
star, we obtained a high-resolution spectrum of \hr\ on UT 5 July 2007
using the HIRES instrument \citep{Vogt:94} on the Keck~I telescope.
The spectrometer slit was $0\farcs86$, giving a resolving power of
$\lambda/\Delta\lambda \approx 55,\!000$. The signal-to-noise ratio of
this spectrum is $\sim$120 per pixel. The spectral region
$\lambda\lambda$6200--6300~\AA\ was used to measure a number of
temperature-sensitive lines mainly of \ion{Fe}{1} and \ion{V}{1}, and
also a few of \ion{Ni}{1}, \ion{Sc}{1}, \ion{Si}{1}, and
\ion{Fe}{2}. As described by \cite{Gray:91} and others, line-depth
ratios (LDRs) of properly selected line pairs are an excellent
diagnostic of effective temperature that allow for \emph{relative}
measurements with a precision as small as a few K, not only in dwarfs
but also in giants \citep{Gray:01}.  The conversion to an
\emph{absolute} temperature scale, however, necessarily depends on an
external color-temperature calibration since LDR variations are
usually compared with corresponding changes in a color index such as
$B\!-\!V$, which is an easier quantity to measure than temperature.
Therefore, absolute temperatures are much less certain. We measured a
total of 26 lines of the primary star\footnote{The secondary is too
faint to affect these measurements, and would not influence them even
if it were brighter \citep[see][]{Gray:01}.} from Table~1 by
\cite{Biazzo:07}, and used their calibrations appropriate for giants
for 16 selected line pairs to derive an average temperature from the
LDRs of $T_{\rm eff} = 4340 \pm 20$~K (formal error).  The scatter of
the individual determinations is only 80~K. The color-temperature
relation on which the above calibrations are based is that of
\cite{Gray:05}, which combines dwarf and giant temperatures obtained
by many different methods. For this work we have preferred to rely on
a more sophisticated color-temperature relation such as that by
\cite{Ramirez:05}, which accounts not only for luminosity class but
also metallicity, and is based on effective temperatures derived
homogeneously by the Infrared Flux Method. The conversion from the
above LDR-based $T_{\rm eff}$ back to an average color for the primary
star was made using the same prescription by \cite{Gray:05}, and gives
$B\!-\!V = 1.328 \pm 0.014$. The \cite{Ramirez:05} calibration for an
adopted metallicity of [Fe/H] $= -0.24$ (\S\ref{sec:evolution}) then
yields $T_{\rm eff} = 4210 \pm 60$~K, in which the error combines
photometric uncertainties and the scatter of the calibration. This
result is in virtually perfect agreement with the temperature
predicted by the \cite{Girardi:00} models, supporting our overall
conclusions from the previous section.

\section{Discussion and concluding remarks}

The example of \hr\ is one of a growing number of binaries in which
the \hip\ intermediate data have been brought to bear on the orbit of
the system, providing complementary information to that afforded by
other types of observations \citep[see, e.g.,][]{Pourbaix:04,
Fekel:05, Fekel:07, Torres:06, Torres:07a, Torres:07b}. In this case
the \hip\ data yield the inclination angle, and allow the absolute
masses to be derived for the first time. Additionally they provide an
estimate of the brightness difference.

Although the inclination we derive for the orbit is consistent with
90\arcdeg\ and allows for the possibility of eclipses, chances are
slim because of the long period compounded by the relatively large
uncertainty in $i$. No photometric variability was found by
\cite{Percy:93}, who, however, apparently did not observe during the
eclipse phases, or by the \hip\ satellite (scatter $\sigma_{H_p} =
0.007$ mag). The latter observations do straddle two of the eclipses,
which can be predicted very accurately from the spectroscopic orbit of
\cite{Scarfe:07} with errors of only 1.0 days for primary eclipse (HJD
$2,\!448,\!652.4$) and 1.4 days for the secondary (HJD
$2,\!448,\!419.3$).  If central, the eclipses would last approximately
18 and 22 days, respectively.  Examination of the epoch photometry
from \hip\ indicates that both of these eclipses were missed by just a
few days. Future eclipses would be expected to occur at Julian dates
$2,\!455,\!254.7 \pm 1.2$ and $2,\!457,\!455.5 \pm 1.4$ for the
primary, and $2,\!455,\!021.6 \pm 1.4$ and $2,\!457,\!222.4 \pm 1.6$
for the secondary. The small separation of only 233 days (10.6\% of a
cycle) between the secondary eclipses and the primary events that
follow is due to a combination of high eccentricity and a longitude of
periastron near zero.

A second implication of the near edge-on orientation of \hr\ is
significantly lower component masses than previously assumed, which in
turn leads to a linear semimajor axis for the relative orbit of $4.63
\pm 0.11$ AU, corresponding to $26.0 \pm 2.4$ mas. The maximum angular
separation subtended by the binary is about 43 mas, also considerably
smaller than has been suggested in the past \citep[$\sim$70
mas;][]{McAlister:76, Halbwachs:81}. About two dozen unsuccessful
attempts to detect the secondary have been made over the past 3
decades with the speckle technique. Some of them were made at phases
in the orbit when the instrumental resolution should have allowed the
observers to resolve the pair, suggesting the large disparity in
brightness ($\Delta V \approx 3$) as the cause of those
non-detections. In principle the expected separations make it a good
target for long-baseline interferometry, except that in most cases
these instruments observe in the near infrared, where the brightness
difference will be even more extreme. From our modeling we expect
$\Delta J \sim 3.8$ mag and $\Delta K \sim 4.1$ mag. Still, some of
these facilities may have the sensitivity required. They could also
resolve the primary itself; the angular diameter should be
approximately 1.6 mas, while that of the secondary is only 0.2 mas.

Double-lined spectroscopic binaries with two giant components are a
relatively rare occurrence, and usually imply the components must have
nearly identical mass.  \hr\ is quite remarkable in that the stars
have attained a large brightness difference that is near the maximum
for this system, yet the secondary is still visible. Evolution
proceeds very rapidly at this stage, and according to the models the
difference in brightness may still increase by another half a
magnitude or so in $V$ over the next 20~Myr, which represents only
$\sim$0.5\% of its present age.
	
\acknowledgements 

We are grateful to an anonymous referee for helpful suggestions on the
original manuscript.  Partial support for this work from NSF grant
AST-0708229 and NASA's MASSIF SIM Key Project (BLF57-04) are
acknowledged. This research has made use of the VizieR service
\citep{Ochsenbein:00} and of the SIMBAD database, both operated at
CDS, Strasbourg, France, as well as of NASA's Astrophysics Data System
Abstract Service. Alex Sozzetti is thanked for help in obtaining and
reducing the Keck spectrum of \hr.


\LongTables
\begin{deluxetable}{lcccccc}
\tabletypesize{\scriptsize}
\tablewidth{0pc}
\tablecaption{\hip\ Intermediate Astrometric Data for \hr.\label{tab:abscissae}}
\tablehead{\colhead{Date} & \colhead{} & \colhead{Orbital} & \colhead{$\Delta v$} & 
\colhead{$\sigma_{\Delta v}$} & \colhead{$O\!-\!C$} & \colhead{Consortium} \\
\colhead{(HJD$-2,\!400,\!000$)} & \colhead{Year} & \colhead{Phase} & 
\colhead{(mas)} & \colhead{(mas)} & \colhead{(mas)} & \colhead{(NDAC/FAST)}}
\startdata
 47862.0382\dotfill  &  1989.9165  &  0.6830  & \phn$+$3.09  &  2.68  &  $+$2.12  &   N  \\
 47899.4398\dotfill  &  1990.0187  &  0.6999  & \phn$-$1.33  &  3.49  &  $-$1.89  &   F  \\
 47899.4398\dotfill  &  1990.0189  &  0.7000  & \phn$-$2.65  &  2.55  &  $-$3.21  &   N  \\
 47926.5048\dotfill  &  1990.0931  &  0.7123  & \phn$+$2.13  &  1.65  &  $+$1.79  &   F  \\
 47926.5048\dotfill  &  1990.0930  &  0.7123  & \phn$+$1.63  &  1.69  &  $+$1.29  &   N  \\
 47985.6022\dotfill  &  1990.2552  &  0.7392  & \phn$-$0.49  &  1.76  &  $-$0.33  &   F  \\
 47985.6022\dotfill  &  1990.2549  &  0.7392  & \phn$-$2.97  &  1.71  &  $-$2.81  &   N  \\
 47999.3721\dotfill  &  1990.2924  &  0.7454  & \phn$-$0.85  &  2.00  &  $-$0.89  &   F  \\
 47999.2991\dotfill  &  1990.2924  &  0.7454  & \phn$+$0.61  &  1.47  &  $+$0.57  &   N  \\
 48053.5022\dotfill  &  1990.4408  &  0.7700  & \phn$-$0.96  &  1.57  &  $-$0.71  &   F  \\
 48053.5022\dotfill  &  1990.4408  &  0.7700  & \phn$-$1.62  &  1.43  &  $-$1.37  &   N  \\
 48091.2325\dotfill  &  1990.5437  &  0.7871  & \phn$+$1.17  &  2.71  &  $+$0.93  &   F  \\
 48091.2325\dotfill  &  1990.5432  &  0.7870  & \phn$+$2.25  &  2.73  &  $+$2.01  &   N  \\
 48115.2294\dotfill  &  1990.6101  &  0.7981  & \phn$+$5.42  &  2.38  &  $+$5.31  &   F  \\
 48115.2294\dotfill  &  1990.6098  &  0.7981  & \phn$+$2.05  &  2.39  &  $+$1.94  &   N  \\
 48144.5225\dotfill  &  1990.6901  &  0.8114  & \phn$+$1.04  &  1.94  &  $+$0.95  &   F  \\
 48144.5225\dotfill  &  1990.6901  &  0.8114  & \phn$+$0.65  &  2.09  &  $+$0.56  &   N  \\
 48181.7780\dotfill  &  1990.7918  &  0.8283  & \phn$+$3.07  &  2.74  &  $+$2.72  &   N  \\
 48182.2163\dotfill  &  1990.7932  &  0.8285  & \phn$-$1.53  &  1.94  &  $-$1.86  &   F  \\
 48182.2163\dotfill  &  1990.7935  &  0.8285  & \phn$-$0.43  &  2.22  &  $-$0.76  &   N  \\
 48201.2823\dotfill  &  1990.8454  &  0.8372  & \phn$-$0.61  &  1.93  &  $+$0.28  &   F  \\
 48201.2823\dotfill  &  1990.8454  &  0.8372  & \phn$+$2.77  &  3.06  &  $+$3.66  &   N  \\
 48243.0304\dotfill  &  1990.9602  &  0.8562  & \phn$-$1.98  &  2.67  &  $-$0.73  &   F  \\
 48243.0304\dotfill  &  1990.9598  &  0.8561  & \phn$-$1.02  &  3.05  &  $+$0.23  &   N  \\
 48266.5891\dotfill  &  1991.0237  &  0.8667  & \phn$-$2.61  &  2.89  &  $-$1.37  &   F  \\
 48266.5525\dotfill  &  1991.0243  &  0.8668  & \phn$-$5.34  &  3.20  &  $-$4.09  &   N  \\
 48267.0273\dotfill  &  1991.0245  &  0.8669  & \phn$-$2.19  &  2.50  &  $-$0.98  &   F  \\
 48266.9908\dotfill  &  1991.0245  &  0.8669  & \phn$+$0.73  &  2.67  &  $+$1.94  &   N  \\
 48294.9690\dotfill  &  1991.1022  &  0.8798  & \phn$-$5.80  &  2.92  &  $-$3.10  &   F  \\
 48294.9690\dotfill  &  1991.1019  &  0.8797  & \phn$-$4.27  &  3.35  &  $-$1.58  &   N  \\
 48295.4073\dotfill  &  1991.1027  &  0.8799  & \phn$-$1.66  &  2.89  &  $+$1.04  &   F  \\
 48295.4073\dotfill  &  1991.1031  &  0.8799  & \phn$-$0.94  &  3.18  &  $+$1.76  &   N  \\
 48327.3666\dotfill  &  1991.1911  &  0.8945  & \phn$+$2.25  &  3.32  &  $+$1.51  &   F  \\
 48327.3666\dotfill  &  1991.1906  &  0.8944  & \phn$-$0.54  &  3.72  &  $-$1.28  &   N  \\
 48327.8050\dotfill  &  1991.1918  &  0.8946  & \phn$+$2.48  &  3.24  &  $+$1.77  &   N  \\
 48344.7526\dotfill  &  1991.2382  &  0.9023  & \phn$+$1.05  &  2.43  &  $+$3.80  &   F  \\
 48344.6795\dotfill  &  1991.2378  &  0.9023  & \phn$-$2.10  &  5.27  &  $+$0.63  &   N  \\
 48384.6744\dotfill  &  1991.3475  &  0.9205  & \phn$+$3.62  &  2.18  &  $-$0.05  &   F  \\
 48384.6744\dotfill  &  1991.3475  &  0.9205  & \phn$+$7.30  &  3.27  &  $+$3.63  &   N  \\
 48395.7049\dotfill  &  1991.3779  &  0.9255  & \phn$+$2.72  &  1.82  &  $+$1.70  &   F  \\
 48395.7049\dotfill  &  1991.3777  &  0.9255  & \phn$+$3.88  &  2.17  &  $+$2.86  &   N  \\
 48436.1381\dotfill  &  1991.4882  &  0.9438  & \phn$+$3.63  &  2.22  &  $-$0.93  &   F  \\
 48436.1016\dotfill  &  1991.4884  &  0.9439  & \phn$+$3.91  &  2.95  &  $-$0.65  &   N  \\
 48455.2042\dotfill  &  1991.5408  &  0.9526  & \phn$+$4.28  &  2.58  &  $-$0.43  &   F  \\
 48455.2042\dotfill  &  1991.5406  &  0.9525  & \phn$+$4.37  &  3.14  &  $-$0.34  &   N  \\
 48488.4784\dotfill  &  1991.6314  &  0.9676  & \phn$+$4.01  &  2.03  &  $+$2.86  &   F  \\
 48488.4419\dotfill  &  1991.6316  &  0.9676  & \phn$+$3.98  &  2.72  &  $+$2.83  &   N  \\
 48519.9995\dotfill  &  1991.7182  &  0.9820  & \phn$+$6.03  &  1.68  &  $+$0.35  &   F  \\
 48519.9995\dotfill  &  1991.7180  &  0.9820  & \phn$+$6.31  &  2.24  &  $+$0.62  &   N  \\
 48543.0468\dotfill  &  1991.7812  &  0.9925  & \phn$-$4.35  &  1.92  &  $-$1.21  &   F  \\
 48543.0468\dotfill  &  1991.7811  &  0.9924  & \phn$-$3.45  &  1.85  &  $-$0.31  &   N  \\
 48586.1097\dotfill  &  1991.8990  &  0.0120  & \phn$-$2.14  &  2.13  &  $-$0.43  &   F  \\
 48586.1097\dotfill  &  1991.8990  &  0.0120  & \phn$+$3.20  &  3.17  &  $+$4.91  &   N  \\
 48606.0524\dotfill  &  1991.9536  &  0.0211  & \phn$-$2.08  &  1.91  &  $+$1.98  &   F  \\
 48606.0524\dotfill  &  1991.9537  &  0.0211  & \phn$-$0.31  &  2.82  &  $+$3.75  &   N  \\
 48640.2398\dotfill  &  1992.0469  &  0.0366  & \phn$-$2.43  &  2.93  &  $-$0.06  &   F  \\
 48640.2033\dotfill  &  1992.0471  &  0.0366  & \phn$+$2.16  &  3.14  &  $+$4.53  &   N  \\
 48669.0580\dotfill  &  1992.1261  &  0.0497  & \phn$-$1.67  &  1.95  &  $-$1.60  &   F  \\
 48669.0580\dotfill  &  1992.1261  &  0.0497  & \phn$-$1.80  &  2.00  &  $-$1.73  &   N  \\
 48727.5346\dotfill  &  1992.2865  &  0.0763  & \phn$-$1.45  &  2.97  &  $-$0.21  &   F  \\
 48727.5346\dotfill  &  1992.2862  &  0.0763  & \phn$-$3.25  &  3.14  &  $-$2.02  &   N  \\
 48728.0459\dotfill  &  1992.2872  &  0.0764  & \phn$+$6.85  &  3.47  &  $+$8.06  &   F  \\
 48728.0094\dotfill  &  1992.2875  &  0.0765  & \phn$+$4.19  &  4.20  &  $+$5.40  &   N  \\
 48739.9896\dotfill  &  1992.3200  &  0.0819  & \phn$+$0.44  &  1.82  &  $+$0.09  &   F  \\
 48739.9896\dotfill  &  1992.3202  &  0.0819  & \phn$+$2.03  &  2.19  &  $+$1.68  &   N  \\
 48780.7880\dotfill  &  1992.4323  &  0.1005  & \phn$-$4.94  &  1.99  &  $-$0.37  &   F  \\
 48780.7880\dotfill  &  1992.4320  &  0.1005  & \phn$-$3.69  &  1.90  &  $+$0.87  &   N  \\
 48781.2263\dotfill  &  1992.4333  &  0.1007  & \phn$-$5.75  &  3.30  &  $-$1.15  &   N  \\
 48795.8728\dotfill  &  1992.4736  &  0.1074  & \phn$-$0.22  &  2.09  &  $+$3.70  &   F  \\
 48795.8728\dotfill  &  1992.4733  &  0.1073  & \phn$-$3.51  &  3.01  &  $+$0.41  &   N  \\
 48832.2517\dotfill  &  1992.5731  &  0.1239  & \phn$+$0.26  &  2.23  &  $+$3.18  &   F  \\
 48832.2517\dotfill  &  1992.5725  &  0.1238  & \phn$-$5.10  &  3.03  &  $-$2.18  &   N  \\
 48926.3036\dotfill  &  1992.8306  &  0.1666  & \phn$-$0.19  &  3.83  &  $+$1.97  &   F  \\
 48926.3036\dotfill  &  1992.8304  &  0.1666  & \phn$-$1.81  &  4.28  &  $+$0.34  &   N  \\
 48944.4931\dotfill  &  1992.8802  &  0.1749  & $+$11.27  &  1.89  &  $+$0.50  &   F  \\
 48944.4931\dotfill  &  1992.8801  &  0.1748  & $+$11.24  &  2.54  &  $+$0.47  &   N  \\
 49025.1768\dotfill  &  1993.1013  &  0.2116  & $+$15.87  &  1.89  &  $+$2.19  &   F  \\
 49025.1768\dotfill  &  1993.1011  &  0.2115  & $+$10.99  &  3.22  &  $-$2.69  &   N  \\
\noalign{\vskip -6pt}
\enddata
\end{deluxetable}


\begin{deluxetable}{lccccc}
\tablewidth{0pc}
\tablecaption{Astrometric orbital solutions for \hr.\label{tab:elements}}
\tablehead{\colhead{\hfil~~~~~~~~~~~Parameter~~~~~~~~~~~~} & \colhead{\cite{Jancart:05}\tablenotemark{a}} & \colhead{This work\tablenotemark{b}}}
\startdata
\noalign{\vskip -2pt}
\sidehead{Adjusted quantities} \\
\noalign{\vskip -8pt}
~~~~$P$ (days)\dotfill                         &  2150 (fixed)                &  2200.77 (fixed) \\
~~~~$a_{\rm phot}$ (mas)\dotfill               &  16.5~$\pm$~2.3\phn          &  9.69~$\pm$~0.85 \\
~~~~$e$\dotfill                                &  0.6 (fixed)                 &  0.6797 (fixed) \\          
~~~~$i$ (deg)\dotfill                          &  46.4~$\pm$~3.5\phn          &  91.4~$\pm$~9.7\phn \\
~~~~$\omega_{\rm A}$ (deg)\dotfill             &  340 (fixed)                 &  9.73 (fixed) \\
~~~~$\Omega_{\rm J2000}$ (deg)\dotfill         &  305.4~$\pm$~6.7\phn\phn     &  50.5~$\pm$~6.6\phn \\
~~~~$T$ (HJD$-$2,400,000)\dotfill              &  50699 (fixed)\tablenotemark{c}               &  50760.4 (fixed) \\
~~~~$\Delta\alpha^*$ (mas)\dotfill             &  \nodata                     &  $+$4.41~$\pm$~0.76\phs \\
~~~~$\Delta\delta$ (mas)\dotfill               &  \nodata                     &  $+$3.77~$\pm$~0.95\phs \\
~~~~$\Delta\mu_{\alpha}^*$ (mas~yr$^{-1}$)\dotfill   &  \nodata               &  $-$3.57~$\pm$~0.91\phs \\
~~~~$\Delta\mu_{\delta}$ (mas~yr$^{-1}$)\dotfill     &  \nodata               &  $-$3.13~$\pm$~1.23\phs \\
~~~~$\Delta\pi_{\rm t}$ (mas)\dotfill          &  \nodata                     &  $+$0.27~$\pm$~0.51\phs \\
\sidehead{Derived quantities} \\						                                         
\noalign{\vskip -8pt}										                                         
~~~~$\mu_{\alpha}^*$ (mas~yr$^{-1}$)\dotfill   &  \nodata                     &  $-$12.46~$\pm$~0.91\phn\phs \\
~~~~$\mu_{\delta}$ (mas~yr$^{-1}$)\dotfill     &  \nodata                     &  $-$31.13~$\pm$~1.23\phn\phs \\
~~~~$\pi_{\rm t}$ (mas)\dotfill                &  \nodata                     &  5.61~$\pm$~0.51 \\
~~~~$M_{\rm A}$ (M$_{\sun}$)\dotfill           &  \nodata                     &  $1.38_{-0.03}^{+0.09}$ \\
~~~~$M_{\rm B}$ (M$_{\sun}$)\dotfill           &  \nodata                     &  $1.36_{-0.02}^{+0.07}$ \\
\noalign{\vskip -6pt}
\enddata

\tablenotetext{a}{$P$, $e$, $\omega_{\rm A}$, and $T$ adopted from
\cite{Christie:36}. Other {\it Hipparcos}-related parameters were not
reported.}

\tablenotetext{b}{$P$, $e$, $\omega_{\rm A}$, and $T$ adopted from
\cite{Scarfe:07}.}

\tablenotetext{c}{Projected forward from the original epoch
$2,\!424,\!290$ (HJD) using the more accurate period from the present
paper, for easier comparison with our new results.}

\end{deluxetable}


\begin{deluxetable}{lccc}
\tablewidth{0pt}
\tablecaption{Comparison of the proper motion and parallax results for \hr.\label{tab:pm}}
\tablehead{\colhead{~~~~~~~~~~Parameter~~~~~~~~~~} & \colhead{\it Hipparcos}  &
\colhead{Tycho-2} & \colhead{This paper} }
\startdata
~~~$\mu_{\alpha}^*$ (mas yr$^{-1}$)\dotfill  & $-8.89$~$\pm$~0.58\phs      &  $-13.3$~$\pm$~1.4\phn\phs  &  $-12.46$~$\pm$~0.91\phn\phs   \\
~~~$\mu_{\delta}$ (mas yr$^{-1}$)\dotfill    & $-28.00$~$\pm$~0.60\phn\phs &  $-30.8$~$\pm$~1.4\phn\phs  &  $-31.13$~$\pm$~1.23\phn\phs    \\
~~~$\pi_{\rm t}$ (mas)\dotfill               & 5.34~$\pm$~0.60             &  \nodata                    & 5.61~$\pm$~0.51 \\
\noalign{\vskip -6pt}
\enddata
\end{deluxetable}


\begin{deluxetable}{lccccc}
\tablewidth{0pt}
\tablecaption{Properties for the components of \hr\ inferred from the models.\label{tab:models}}
\tablehead{\colhead{~~~~~~~~~~Property~~~~~~~~~~} & \colhead{$Z = 0.004$}  &
\colhead{$Z = 0.008$} & \colhead{$Z = 0.011$} & \colhead{$Z = 0.019$} & 
\colhead{$Z = 0.030$} }
\startdata

Log age (yr)\dotfill                & 9.40 & 9.45 & 9.49 & 9.55 & 9.60 \\
$(B\!-\!V)_{\rm A}$ (mag)\dotfill   & 1.07 & 1.25 & 1.35 & 1.50 & 1.58 \\
$(B\!-\!V)_{\rm B}$ (mag)\dotfill   & 0.75 & 0.87 & 0.92 & 1.02 & 1.12 \\
$(B\!-\!V)_{\rm tot}$ (mag)\dotfill & 1.05 & 1.22 & 1.32 & 1.47 & 1.54 \\
$\Delta J$ (mag)\dotfill            & 3.48 & 3.64 & 3.80 & 3.96 & 4.11 \\
$\Delta H$ (mag)\dotfill            & 3.66 & 3.88 & 4.05 & 4.29 & 4.43 \\
$\Delta K$ (mag)\dotfill            & 3.68 & 3.90 & 4.08 & 4.33 & 4.54 \\
$T_{\rm eff}^{\rm A}$ (K)\dotfill   & 4590 & 4311 & 4211 & 3947 & 3699 \\
$T_{\rm eff}^{\rm B}$ (K)\dotfill   & 5312 & 5079 & 5010 & 4819 & 4649 \\
$R_{\rm A}$ (R$_{\sun}$)\dotfill    & 23.1 & 29.3 & 30.9 & 42.1 & 56.7 \\
$R_{\rm B}$ (R$_{\sun}$)\dotfill    & 3.84 & 4.33 & 4.20 & 4.98 & 5.90 \\
\noalign{\vskip -6pt}
\enddata 

 \tablecomments{The models used are those of \cite{Girardi:00}, and
the observational constraints are the measured masses and absolute
magnitudes of the components (see text). The best-fitting model for $Z
= 0.011$ satisfies the additional constraint that the combined colors
of the components match the observed $B\!-\!V$ color.}

\end{deluxetable}

\end{document}